\documentclass{article}
\usepackage{dcase2021_techrep,amsmath,graphicx,url,times, booktabs, tabularx}
\usepackage{multirow}
\usepackage{algorithm}
\usepackage{algpseudocode}
\usepackage{enumitem}
\usepackage{siunitx}
\usepackage[numbers, sort&compress,square]{natbib}
\usepackage{placeins}
\usepackage{multicol, multirow}
\usepackage{booktabs}
\usepackage[nameinlink]{cleveref}
\usepackage{lipsum}
\usepackage{caption}
\usepackage{subcaption}
\usepackage{comment}

\newcolumntype{R}{>{\raggedleft\arraybackslash}X}
\newcolumntype{L}{>{\raggedright\arraybackslash}X}
\newcolumntype{C}{>{\centering\arraybackslash}X}
\newcommand{\etal}{~\emph{et~al.} }

\usepackage[usenames, dvipsnames]{color} 
\definecolor{purple}{rgb}{0.5, 0.0, 0.5}
\definecolor{orange}{rgb}{1, 0.65, 0}
\definecolor{lightgreen}{rgb}{0.68, 1, 0.18}
\definecolor{darkgreen}{rgb}{0.09, 0.32, 0.24}
\definecolor{darkred}{rgb}{0.6, 0, 0}
\definecolor{brown}{rgb}{0.64, 0.16, 0.16}
\iftrue 
  \newcommand{\tho}[1]{\noindent}
  \newcommand{\karn}[1]{\noindent}
  \newcommand{\done}[1]{\noindent}
  \newcommand{\todo}[1]{\noindent}
\else
  \newcommand{\tho}[1]{\textcolor{blue}{\bf [V: #1]}}
  \newcommand{\karn}[1]{\textcolor{orange}{\bf [T: #1]}}
  \newcommand{\done}[1]{\textcolor{darkgreen}{\bf [Done: #1]}}
  \newcommand{\todo}[1]{\textcolor{red}{\bf [Todo: #1]}}
\fi
\usepackage[normalem]{ulem} 

\title{DCASE 2021 TASK 3: SPECTROTEMPORALLY-ALIGNED FEATURES\\FOR POLYPHONIC SOUND EVENT LOCALIZATION AND DETECTION}

\name{
    Thi Ngoc Tho Nguyen$^{1}
        \thanks{This research was supported by the Singapore Ministry of Education Academic Research Fund Tier-2, under research grant MOE2017-T2-2-060.}$,
	Karn Watcharasupat$^{1}
	    \thanks{K. Watcharasupat was supported by the CN Yang Scholars Programme, Nanyang Technological University.}$,
} \secondlinename{	  
	Ngoc Khanh Nguyen,
    Douglas L. Jones$^{2}$, 
    Woon Seng Gan$^{1}$
}

\address{
    $^1$ School of Electrical and Electronic Engineering, Nanyang Technological University, Singapore.\\          
    $^2$ Dept. of Electrical and Computer Engineering, University of Illinois at Urbana-Champaign, IL, USA. \\
    \{nguyenth003, karn001\}@e.ntu.edu.sg, ngockhanh5794@gmail.com,\\ dl-jones@illinois.edu, ewsgan@ntu.edu.sg
}

\begin{document}

\ninept
\maketitle

\begin{sloppy}

\begin{abstract}
Sound event localization and detection consists of two subtasks which are sound event detection and direction-of-arrival estimation. While sound event detection mainly relies on time-frequency patterns to distinguish different sound classes, direction-of-arrival estimation uses magnitude or phase differences between microphones to estimate source directions. Therefore, it is often difficult to jointly optimize these two subtasks simultaneously. We propose a novel feature called spatial cue-augmented log-spectrogram (SALSA) with exact time-frequency mapping between the signal power and the source direction-of-arrival. The feature includes multichannel log-spectrograms stacked along with the estimated direct-to-reverberant ratio and a normalized version of the principal eigenvector of the spatial covariance matrix at each time-frequency bin on the spectrograms. Experimental results on the DCASE 2021 dataset for sound event localization and detection with directional interference showed that the deep learning-based models trained on this new feature outperformed the DCASE challenge baseline by a large margin. We combined several models with slightly different architectures that were trained on the new feature to further improve the system performances for the DCASE sound event localization and detection challenge.  
\end{abstract}

\begin{keywords}
DCASE, deep learning, spatial audio, feature extraction, sound event localization and detection. 
\end{keywords}

\section{Introduction}
\label{sec:intro}

Sound event localization and detection (SELD) has many applications in urban sound sensing~\cite{Salamon2017Cnn}, wildlife monitoring~\cite{Stowell2016bird}, and surveillance~\cite{Foggia2016Surveillance}. SELD is the problem of recognizing the sound class, as well as estimating the corresponding direction of arrival (DOA), onset, offset of the detected sound event~\cite{adavanne2019seld}. Polyphonic SELD refers to cases where there are multiple sound events overlapping in time. In 2019, the Challenge on Detection and Classification of Acoustic Scenes and Events (DCASE) introduces a polyphonic SELD task with only stationary sound sources~\cite{adavanne2019dcasedataset}. The 2020 rendition sees an introduction of moving sound sources~\cite{politis2020dcasedataset}. In 2021, the SELD task additionally introduces unknown directional interferences that further complicate the task~\cite{politis2021dcasedataset}. 

SELD is an emerging topic in audio processing. It consists of two subtasks, which are sound event detection (SED) and DOA estimation (DOAE). These two subtasks are mature research topics, and there exists a large body of effective algorithms for SED and DOAE~\cite{cakir2017convolutional, mohan2008localization}. Over the past few years, the majority of the methods proposed for SELD have focused on jointly optimizing SED and DOAE in one network. Hirvonen first formulated SELD as a multi-class classification task where the number of output classes is equal to the number of DOAs multiplied by the number of sound classes~\cite{hirvonen2015classification}. In 2018, Adavanne\etal pioneered a seminal work that used a single-input multiple-output convolutional recurrent neural network (CRNN) model called SELDnet, which jointly detects sound events and estimates the corresponding DOAs~\cite{adavanne2019seld}.  

Because sound event detection mainly relies on time-frequency (TF) patterns to distinguish different sound classes while direction-of-arrival estimation relies on magnitude or phase differences between microphones to estimate source directions, it is often difficult to jointly optimize these two subtasks in a single network. To remedy this problem, Cao\etal proposed a two-stage strategy by training separate SED and DOA models~\cite{cao2019polyphonic}, then using the SED outputs as masks to select DOA outputs. This training scheme significantly improved the SELD performance over the jointly-trained SELDnet. Cao\etal later proposed an end-to-end SELD network called Event Independent Network (EIN)~\cite{cao2020ein, cao2021ienv2} that used soft parameter sharing between the SED and DOAE encoder branches, and segregated the SELD output into event-independent tracks. The second version of EIN that used multi-head self-attention (MHSA) to decode the SELD output is currently the state-of-the-art solution for on DCASE 2020 evaluation set using a single model~\cite{cao2021ienv2}. In another research direction, Shimada\etal proposed a new output format for SELD called activity-coupled Cartesian DOA (ACCDOA) that required only one loss function to optimize the SELD network~\cite{shimada2021accdoa}. The authors also proposed to use a densely connected multi-dilated DenseNet (RD3Net) instead of CRNN to achieve a better SELD performance. The RD3Net with ACCDOA outputs is currently the state-of-the-art solution on the DCASE 2020 test set using a single model. The top-ranked solution for DCASE 2020 SELD challenge synthesized a larger dataset from the provided data using four different data augmentation methods and combined many SELDnet-like models with more complex sub-networks into an ensemble~\cite{du2020dcasetop, wang2021fouraug}.      

When SELDnet was first introduced, it was trained on multichannel magnitude and phase spectrograms~\cite{adavanne2019seld}. Subsequently, different features such as multichannel log-mel spectrograms and intensity vectors (IV) for the first-order ambisonics format (FOA) and generalized cross-correlation with phase transform (GCC-PHAT) for the microphone array (MIC) format were shown to be more effective for the SELD task~\cite{politis2020dcasedataset, politis2021dcasedataset, cao2019polyphonic, cao2021ienv2, shimada2021accdoa, du2020dcasetop, wang2021fouraug, wang2021ensem}. The advantages of the log-mel spectrograms over the linear magnitude spectrograms for deep learning-based SED are lower dimensions and more emphasis on the lower frequency bands where signal contents are mostly populated. 

However, combining IV or GCC-PHAT features with log-mel spectrograms is not trivial and the implicit DOA information stored in the former features are often compromised. In order to stack the IVs with log-mel spectrograms, frequency band compression on the IVs is required. In practice, the IVs are often passed through the mel filters which merge DOA cues in different narrow bands into one mel band, making it more difficult to resolve different DOAs in multi-source scenarios. Nonetheless, although the resolution of the DOA cues is reduced, the corresponding frequency mapping between log-mel spectrograms and the IVs are preserved. This frequency correspondence is crucial for algorithms to associate sound classes and DOAs of multiple sound events, where signals of different sound sources are often distributed differently along the frequency dimension. This frequency correspondence, however, has no counterpart for GCC-PHAT features since the time lags dimension of the GCC-PHAT features does not have a linear one-to-one mapping with the mel bands of the log-mel spectrograms. As a result, all of the DOA information is aggregated at the frame level, making it difficult to assign correct DOAs to different sound events. In addition, GCC-PHAT features are noisy when there are multiple sound sources. In order to solve SELD more effectively in multi-source scenarios with interferences, a better feature is needed for both audio formats. 

In this work, we proposed a novel feature for SELD task called Spatial Cue-Augmented Log-Spectrogram (SALSA) with exact spectrotemporal mapping between the signal power and the source direction-of-arrival. The feature includes multichannel log-spectrograms stacked along with the estimated direct-to-reverberant ratio and a normalized version of the principal eigenvector of the spatial covariance matrix at each TF bin on the spectrograms. The principal eigenvector is normalized such that it represents the inter-channel intensity difference (IID) for the FOA format, or inter-channel phase difference (IPD) for the microphone array format. We evaluated the effectiveness of the proposed feature using the DCASE 2021 SELD dataset with FOA format. Experimental results showed that the deep learning-based models trained with SALSA feature outperformed the DCASE 2021 challenge baseline model that was trained with log-mel spectrograms and IV features by a large margin. We further combined several SELD models with slightly different architectures into ensembles to maximize the performance of our submitted systems to the challenge. The rest of our paper is organized as follows. \Cref{sec:proposed} describes our proposed method with a brief introduction to SALSA features. \Cref{sec:expt} presents the experimental results and discussions, where our submission strategies are elaborated. Finally, we conclude the paper in \Cref{sec:concl}.

\section{The proposed method} \label{sec:proposed}

\begin{figure}[tb]
\centering
\includegraphics[width=0.4\textwidth]{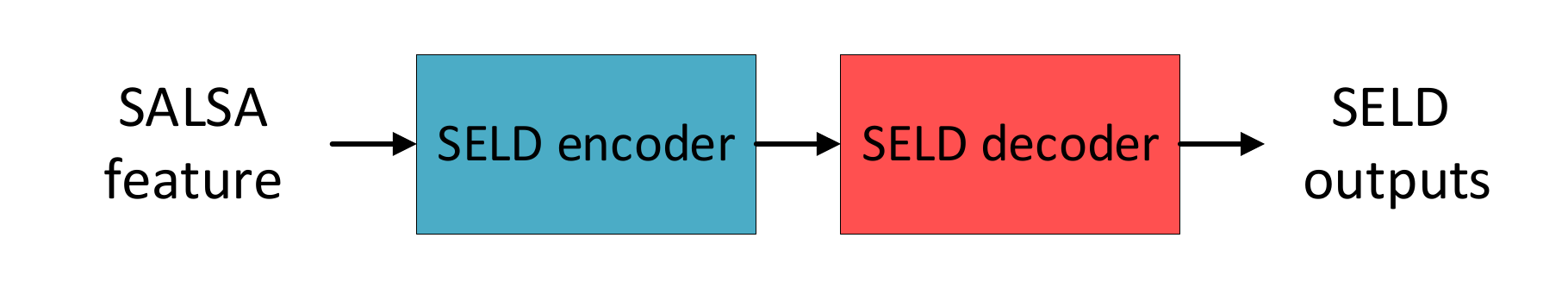}
\caption{SELD network architecture.}
\label{fig1:seld_framework}
\end{figure} 

Figure~\ref{fig1:seld_framework} shows the block diagram of our SELD system. The SELD encoder is a convolutional neural network (CNN) that learns spatial and spectrotemporal representation from SALSA features. The SELD decoder consists of a temporal network and fully connected (FC) layers to decode SELD output sequences. Popular choices for the temporal network are long short-term memory (LSTM), gated recurrent unit (GRU), and MHSA with positional encoding. We train different SELD models with different temporal networks and combine them into ensembles. 

\subsection{Spatial cue-augmented log-spectrogram features (SALSA)}

\begin{figure}[tb]
\centering

\includegraphics[width=\columnwidth]{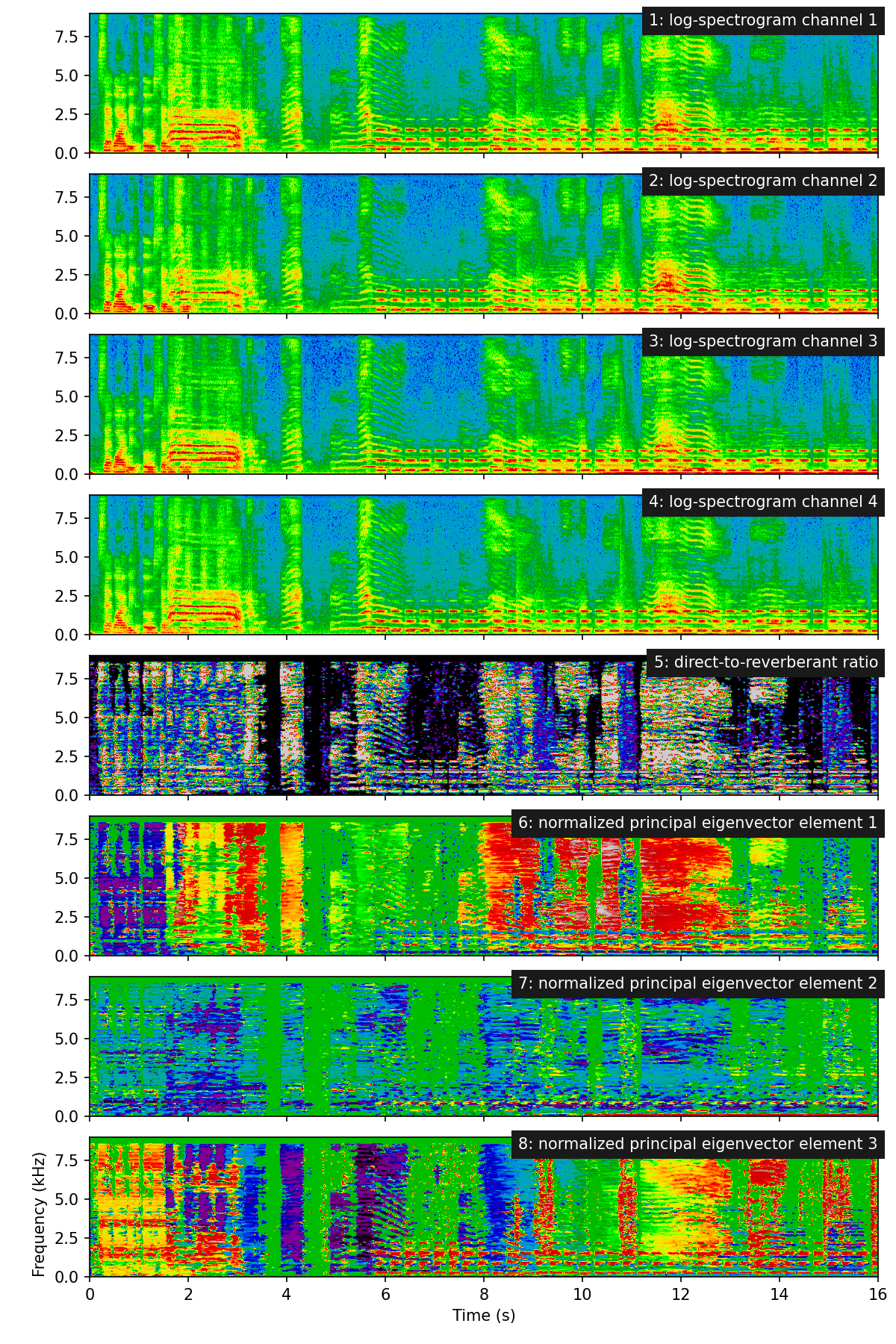}
\captionsetup{belowskip=0pt}
\vspace{-0.5cm}
\caption{SALSA features of a 16-second audio segment with FOA inputs in a multi-source scenario. The horizontal axis represents time in seconds, and the vertical axis represents frequency in kHz.}
\vspace{-0.3cm}
\label{fig2:salsa_feature}
\end{figure} 

\Cref{fig2:salsa_feature} illustrates SALSA features of a 16-second audio segment with four-channel inputs in multi-source cases. The first four channels are log spectrograms. The fifth channel is the estimated direct-to-reverberant ratios (DRR). The last three channels are a normalized version of the principal eigenvectors of the spatial covariance matrix at each time-frequency bin on the spectrograms. The DRR and the eigenvectors are only computed for TF bins whose energy mainly comes from a single sound source. These bins are called single-source (SS) TF bins. In order to find these SS TF bins, we apply a magnitude test and a coherence test on each TF bin~\cite{tho2014robust}. The magnitude test finds all TF bins whose powers are higher than a noise threshold to mitigate the effect of background noise. The coherence test finds TF bins whose spatial covariance matrices are approximately rank-\num{1}. We use DRR which is the ratio between the two largest eigenvalues of the covariance matrix as the criterion for the coherence test. As shown in~\cite{tho2014robust}, the principal eigenvector of the covariance matrix at each SS TF bin is a scaled version of the theoretical steering vector of the corresponding dominant sound source at that particular TF bin. We normalize the principal eigenvectors and remove the redundant element corresponding to the reference microphone. For the FOA format, this normalized eigenvector corresponds to the IID. The SALSA features are real-valued for both FOA and MIC formats. We can see in \Cref{fig2:salsa_feature} that the last four spatial-feature channels are visually discriminant for different sources originating from different directions. 

\subsection{Network architecture}

For the SELD encoder, we use the ResNet22 network, which was adopted for audio tagging application~\cite{kong2019panns}, without the last global pooling and fully connected layer. Unlike the ResNet18 network used in image applications~\cite{resnet}, the first convolution layer of the ResNet22 is a double convolution with kernel size $(3, 3)$ and stride \num{1}. The downsampling factor of ResNet22 is \num{16} on both time and frequency dimensions. The output embedding of the ResNet22 encoder is average-pooled along the frequency dimension before fed into the SELD decoder. For the SELD decoder, we use \num{2} layers of either of the three different temporal networks: bidirectional LSTM, bidirectional GRU, and MHSA with positional encoding. The hidden size of the LSTM and GRU is \num{256}. The dimension of the feedforward network of the MHSA is \num{1024}. The SELD problem is formulated as a multi-label multi-class classification task for sound classes and a regression task for DOAs. We employ \num{4} FC layers to produce the SELD output. The first FC layer, which is followed by a sigmoid activation, produces the posterior probabilities of the sound classes. The remaining three FC layers produce Cartesian coordinates of the DOA on a unit sphere. We call this output format \emph{SEDXYZ}. We also use the newly proposed ACCDOA output format in our experiments. When the ACCDOA format is used, the contribution of the classification loss is set to zero. 

\section{Experimental results and discussions} \label{sec:expt}

\begin{table*}[t!]
    \centering
    \noindent\begin{tabular}{ cr rrrrr rrrrr }
    \toprule 
        \multirow{2}[2]{*}{System} & 
        \multirow{2}[2]{*}{\# Params.} &  
        \multicolumn{5}{c}{Validation} &
        \multicolumn{5}{c}{Test}
    \\ \cmidrule(lr){3-7}\cmidrule(lr){8-12}
        & & 
        $\text{ER}_{\le \SI{20}{\degree}}$ &
        $\text{F}_{\le \SI{20}{\degree}}$ &
        $\text{LE}_\text{CD}$&
        $\text{LR}_\text{CD}$ &
        $\mathcal{D}_\text{SELD}$ &
        $\text{ER}_{\le \SI{20}{\degree}}$ &
        $\text{F}_{\le \SI{20}{\degree}}$ &
        $\text{LE}_\text{CD}$&
        $\text{LR}_\text{CD}$ &
        $\mathcal{D}_\text{SELD}$ 
    \\ \midrule
        A & 112.2M & 0.347 & 0.756 & 13.4 & 0.783 & 0.221 
                   & 0.378 & 0.740 & 11.4 & 0.756 & \bf{0.236}\\
        B & 83.9M  & 0.337 & \bf{0.762} & 13.5 & 0.785 & \bf{0.216} 
                   & 0.376 & 0.738 & \bf{11.2} & 0.750 & 0.238\\
        C & 107.8M & \bf{0.334} & 0.760 & \bf{13.2} & 0.775 & 0.218 
                   & \bf{0.372} & 0.737 & \bf{11.2} & 0.741 & 0.239\\ 
        D & 112.2M & 0.363 & 0.749 & 13.8 & \bf{0.801} & 0.222 
                   & 0.389 & \bf{0.741} & 12.1 & \bf{0.779} & 0.239\\
    \bottomrule
    \end{tabular}
    
    \caption{Evaluation results for submitted systems}
    \label{tab:eval}
\end{table*}

We evaluated our proposed SALSA features using the {TAU-NIGENS} Spatial Sound Events 2021 Dataset~\cite{politis2021dcasedataset}. We compared the performance of different models trained on this new feature against the challenge baseline.

\subsection{Dataset}

We used only the FOA subset of the dataset for our experiments. The development split of the dataset consists of \num{400}, \num{100}, and \num{100} one-minute audio recordings for the train, validation, and test splits respectively. There are \num{12} target sound classes. 
The ground truth metadata provided by the 2021 dataset only includes the labels for sound events belonging to the target classes. In other words, all directional interferences are unlabelled. The azimuth and elevation ranges are $[\SI{-180}{\degree}, \SI{180}{\degree})$ and $[\SI{-45}{\degree}, \SI{45}{\degree}]$, respectively. During development stage, the validation set was used for model selection while the test set was used for evaluation. During evaluation stage, all development data were used for training evaluation models. 

\subsection{Evaluation metrics}

To evaluate the SELD performance, we used the official evaluation metrics~\cite{politis2020overview} that were newly introduced in this year DCASE challenge. The new metrics not only take into account the joint dependencies between SED and DOAE but also penalize systems that cannot resolve the overlapping of multiple instances of the same class~\cite{politis2020overview}. A sound event was considered a correct detection only if it has correct class prediction and its estimated DOA is also less than $T\si{\degree}$ away from the DOA ground truth, where $T=\SI{20}{\degree}$ for this competition. The DOAE metrics are also class-dependent, that is, the detected sound class will also have to be correct in order for the corresponding localization predictions to count.

Similar to DCASE 2020, the DCASE 2021 SELD task adopted four evaluation metrics: location-dependent error rate ($\text{ER}_{\le T\si{\degree}}$) and F1 score ($\text{F}_{\le T\si{\degree}}$) for SED; and class-dependent localization error ($\text{LE}_\text{CD}$), and localization recall ($\text{LR}_\text{CD}$) for DOAE. We also reported an aggregated SELD metric which was computed as
\begin{equation}
    \mathcal{D}_\text{SELD} = \dfrac{1}{4}\left[\text{ER}_{\le T\si{\degree}} + (1-\text{F}_{\le T\si{\degree}}) + \dfrac{\text{LE}_\text{CD}}{\SI{180}{\degree}} + (1-\text{LR}_\text{CD})\right].
\end{equation}
A good SELD system should have low $\text{ER}_{\le T\si{\degree}}$, high $\text{F}_{\le T\si{\degree}}$, low $\text{LE}_\text{CD}$, high $\text{LR}_\text{CD}$, and low aggregated metric $\mathcal{D}_\text{SELD}$. 

\subsection{Hyperparameters and training procedure}

We used sampling rate of \SI{24}{\kilo\hertz}, window length of \num{512} samples, hop length of \SI{300}{samples}, Hann window, and \num{512} FFT points. As a result, the input frame rate of SALSA features was \SI{80}{fps}. Since the model temporally downsampled the input by a factor of \num{16}, we temporally upsampled the final outputs by a factor of \num{2} to match the label frame rate of \SI{10}{fps}. The loss weights for SED and DOAE outputs were set to \num{0.7} and \num{0.3} respectively. Adam optimizer was used to train all the models. Learning rate was initially set to \num{0.003} and gradually decreased to \num{e-4}. The maximum number of training epochs was \num{60}. A threshold of \num{0.3} was used to binarize active class predictions in the SED outputs. 

\subsection{Experimental settings}

We trained several SELD models on the new SALSA features. We compared our models with the DCASE 2021 challenge baseline \cite{adavanne2019seld, shimada2021accdoa} that was trained on log-mel spectrograms and IV features. Since the provided dataset is relatively small, we employed several data augmentation techniques. First, we extended the spatial augmentation technique~\cite{Mazzon2019} that randomly swaps and negates the X, Y, and Z channels of the FOA format to our SALSA feature. Using the provided theoretical steering vector for the FOA format, the last three channels of the SALSA features correspond to Y, Z, and X responses. Therefore, we swapped and negated the last three spatial channels accordingly. Secondly, we applied random cutout~\cite{zhong2017random} and SpecAugment~\cite{park2019specaugment} on all the channels of the SALSA features. For SpecAugment, we only applied time masking and frequency masking. Random cutout produces a rectangular mask on the spectrograms while SpecAugment produces a cross-shaped mask. 
Lastly, we also randomly removed selected SS TF bins on the last 4 channels of the SALSA feature. 

We used different input lengths, e.g. \num{4} seconds, \num{8} seconds, etc., to train different SELD models. We experimented with three different SELD decoders: bidirectional LSTM, bidirectional GRU, and MHSA with positional encoding. We train the majority of these models using the SEDXYZ output format and some models with the ACCDOA output format. Since these two output formats are both class-wise format, they can be easily aggregated into ensembles using mean operation. The disadvantage of the class-wise output format is that they cannot resolve overlapping same-class events, which accounts for \SI{10.45}{\percent} of the total frames in the DCASE 2021 SELD dataset. We chose to use class-wise outputs for the ease of ensemble. To further improve the performance of each trained model, we applied test-time augmentation (TTA) during inference. We adapted the 16-pattern spatial augmentation technique from \cite{Mazzon2019} for TTA, similar to the spatial augmentation technique that was employed during training. We augmented all the channels of the SALSA features (except for the DRR channel), estimated the output, reversed the outputs accordingly, and computed the mean of all the $16$ outputs. 

We combined different SELD models into 4 SELD ensembles. We also train several CRNN models for SED only. We experimented with different combinations of CNN architectures, such as VGG and ResNet, and recurrent neural network (RNN) architectures, such as bidirectional GRU and bidirectional LSTM. We combined SED models with various CNN-RNN combinations into an SED ensemble. This SED ensemble was then combined with the 4 SELD ensembles to form 4 submission systems that were submitted to the challenge. When the SED ensemble and an SELD ensemble were combined, only the SED outputs were averaged, the DOA outputs of the SELD ensemble were kept intact. All four systems used all six folds of the development dataset for training.

\subsection{Experimental results}

\Cref{tab:eval} shows the performances on the validation and test splits of our four submitted systems. System D is similar to System A except for lower SED threshold on common classes such as \emph{foot step}, and \emph{alarm}. As a result, System D scored higher in $\text{LR}_\text{CD}$ at the expense of higher $\text{ER}_{\le \SI{20}{\degree}}$, and $\text{LE}_\text{CD}$ on both validation and test splits. 

\begin{table}[th]
    \centering
    \noindent\begin{tabularx}{\columnwidth}{Lrrrrr}
    \toprule 
        \multirow{2}[2]{*}{Model} & 
        \multirow{2}[2]{*}{\# Params.} &  
        \multicolumn{4}{c}{Test} 
    \\ \cmidrule(lr){3-6}
        & & 
        $\text{ER}_{\le \SI{20}{\degree}}$ &
        $\text{F}_{\le \SI{20}{\degree}}$ &
        $\text{LE}_\text{CD}$&
        $\text{LR}_\text{CD}$ 
    \\ \midrule
        \makebox[4em][l]{Baseline}~LM/IV      
                    & 0.5M      & 0.690 & 0.339 & 24.1 & 0.439 \\
        \makebox[4em][l]{GRU}~LM/IV      
                    & 14.2M     & 0.650 & 0.483 & 22.0 & 0.626 \\
    \midrule
        \makebox[4em][l]{GRU}~w/o~TTA        
                    & 14.2M     & 0.426 & 0.686 & 12.1 & 0.683 \\
        \makebox[4em][l]{}~w/\hphantom{o}~TTA         
                    & 14.2M     & 0.404 & 0.702 & \bf{11.0} & 0.674 \\
        \makebox[4em][l]{LSTM}~w/o~TTA        
                    & 15.0M     & 0.428 & 0.685 & 11.9 & 0.697 \\
        \makebox[4em][l]{}~w/\hphantom{o}~TTA         
                    & 15.0M     & 0.410 & 0.695 & \bf{11.0} & 0.691 \\
        \makebox[4em][l]{MHSA}~w/o~TTA        
                    & 16.1M     & 0.498 & 0.673 & 14.4 & \bf{0.761} \\
        \makebox[4em][l]{}~w/\hphantom{o}~TTA         
                    & 16.1M     & 0.450 & 0.700 & 13.1 & 0.759 \\
    \midrule
        A~w/o~SED ensem.  
                    & 73.6M     & \bf{0.377} & 0.734 & \bf{11.0} & 0.740 \\
        A~w/\hphantom{o}~SED ensem.    
                    & 112.2M    & 0.378 & \bf{0.740} & 11.4 & 0.756 \\
    \bottomrule
    \end{tabularx}
    \caption{Evaluation results for ablation studies. Unless `LM/IV' is indicated, the models are trained on SALSA features. LM/IV stands for log-mel spectrogram and intensity vector features.}
    \label{tab:sys}
\end{table}

\Cref{tab:sys} compares the performance of different SELD models. The challenge baseline, denoted as Baseline LM/IV, and our model GRU LM/IV were trained on the log-mel spectrograms and IV feature. The difference between our SELD model and the baseline CRNN model were that our model had more parameters than the baseline, and we used $128$ mel bands instead of $64$ mel bands. The SELD performance improved with bigger model, GRU LM/IV, especially the $\text{LR}_\text{CD}$ increased from $0.439$ to $0.626$. The rest of our models in \Cref{tab:sys} were trained with SALSA features. The model GRU LM/IV and GRU w/o TTA had almost identical network architectures and similar number of parameters, except that the number of inputs channel to GRU LM/IV was 7, while the number of input channels to GRU w/o TTA was 8. The GRU w/o TTA model outperformed GRU LM/IV by a large margin across all the evaluation metrics. This result demonstrated the effectiveness of our new proposed feature over the conventional log-mel spectrogram and intensity vector feature for SELD using FOA format. When comparing the performance of different SELD decoders, LSTM and GRU achieved similar scores. MHSA scored higher on $\text{LR}_\text{CD}$ than RNN-based decoders but lower on other metrics. TTA improved the performance especially for SED metrics and DOA error. Combining several SELD models into an ensemble boosted the final performance. Interestingly, by combining with an SED ensemble, the performance  of the final ensemble was increased slightly.

\section{CONCLUSION} \label{sec:concl}
In conclusion, we presented a novel spectrotemporally-aligned feature, SALSA, for training a joint end-to-end SELD network. Experimental results showed our networks trained on the new feature outperformed the DCASE 2021 Task 3 baseline system by a large margin, demonstrating the effectiveness of SALSA feature in enabling the deep models to learn useful spatial and spectrotemporal information for SELD task. 

\section{REFERENCES}

\renewcommand{\bibsection}{}
\bibliographystyle{IEEEtran}
\bibliography{refs}

\end{sloppy}
\end{document}